\documentclass{ws-procs9x6}

\begin{document}

\title{Heavy quark production at HERA and its relevance for the LHC}

\author{M. Wing}

\address{Department of Physics and Astronomy\\
University College London\\
Gower Street\\
London WC1E 6BT\\
UK\\
E-mail: mw@hep.ucl.ac.uk}

\maketitle

\abstracts{
The import of HERA data on heavy quark production for LHC experiments is discussed. 
Knowlegde of all aspects of the beauty and charm production process, viz. the parton 
density functions of colliding hadrons, the hard scatter, and the fragmentation of 
the quarks into hadrons, can aid LHC experimentation. This short write-up concentrates 
on possible influences HERA data can have and on the current status (and history) of 
beauty production from 
both HERA and Tevatron experiments. In general, next-to-leading order QCD gives a 
reasonable description of beauty production although some regions of phase space 
such as low $p_T$ show indications of differences.
}

\section{Why study heavy quark production?}

The measurement of heavy quarks can give insights into many physical phenomena such 
as: new particles which are expected to decay predominantly to beauty (and charm); 
precise measurements of electroweak parameters; and, the subject of this paper, a 
deeper understanding of the strong force of nature. The strong force as described 
within perturbative Quantum Chromodynamics (QCD) should be able to give a precise 
description of heavy quark production. This postulate is tested here. 
The measurement of heavy quark production also yields valuable information on the 
structure of colliding hadrons. The production of a pair of heavy quarks in a generic 
hadron collision is directly sensitive to the gluon content of the hadron. Most 
information on the structure of a hadron comes from inclusive deep inelastic 
scattering where the gluon content is determined in the evolution of the QCD 
equations. Therefore measurement of such a process provides complimentary information 
to that from inclusive measurements. 

As well as understanding for its own sake, knowledge of the structure of hadrons 
will be important at future colliders such as the LHC and International Linear 
Collider where hadronic photons will provide a large cross section in both $e^+e^-$ and 
$\gamma \gamma$ modes. Heavy quarks will be copiously produced at future colliders 
as a background to the more exotic processes expected. Therefore a precise description 
of their production properties within QCD will aid in the discovery of physics 
beyond the Standard Model. 

\section{Information needed by the LHC experiments}

Information needed by the LHC which can be provided by the HERA experiments is the 
following:

\begin{itemize}

\item the state of the description of heavy quark production data by theoretical 
      predictions. The production of beauty in the hard scattering process is 
      discussed here in detail. Information on heavy quarks produced in the splitting of 
      a gluon outgoing from the hard sub-process is also important for the LHC, but the 
      information from HERA is currently limited;

\item the gluon and heavy quark proton parton density functions;

\item details of fragmentation in a hadronic environment;

\item the effect of the underlying event in heavy quark processes. This information is 
      limited at HERA but may be studied in the future;

\item HERA results can provide general information on event and jet topologies which will 
      be useful for designing algorithms or triggers at the LHC experiments.

\end{itemize}

Designing effective triggers for $b$ physics is particularly acute for the LHCb 
experiment\cite{misc:brook:private}. Large backgrounds are expected although event 
topologies should be different to the signal $b$ physics. For example minimum bias events 
will have a smaller track multiplicity and a lower transverse momentum for the highest 
$p_T$ track. Therefore using Monte Carlo simulation, cuts can be found to be able to reduce 
the rate of minimum bias whilst triggering efficiently on $b$ events. Such trigger designs  
require reliable Monte Carlo simulation of the event topologies of both classes of events.

Measurements of the proton structure function at HERA will constrain the parton densities 
in a large region of the kinematic plane where $B$ mesons will be produced within the 
acceptance of the LHCb detector. According to Monte Carlo simulations, these events are 
produced predominantly with a $b$ quark in the proton. However, this is just a model 
({\sc Pythia}\cite{cpc:135:238}) and at NLO some of the events will be summed into the 
gluon distribution of the proton. Nevertheless, measuring all flavours in the proton at 
HERA is one of the goals of the experiments and recent results on the beauty contribution 
to the proton structure function\cite{h1-f2b} shed some light on the issue. 

%
%

\section{Open beauty production}

The production of open beauty and its description by QCD has been of great interest in the 
last 10--15 years. The difference between the rates observed by the Tevatron 
experiments\cite{b-tevatron1} and NLO QCD predictions led to a mini crisis with many 
explanations put forward. Several measurements were performed in different decay channels and 
then extrapolated to the quark level to facilitate a comparison with QCD and between themselves. 
The NLO QCD prediction was found to be a factor of 2--3 below the data for all measurements. 
These results were extrapolated to the $b$-quark level using Monte Carlo models which may or 
may not give a good estimate of this extrapolation. To facilitate a particular comparison, 
an extrapolation can be useful, but should always be treated with caution and the procedure 
clearly stated and values of extrapolation factors given. Initial measurements in terms of 
measured quantities should also always be given.

The CDF collaboration also published measurements of $B$ meson cross sections. They were also 
found to be significantly above NLO calculations, but allowed for phenomenological study. Work 
on the fragmentation function was performed by Cacciari and Nason\cite{prl:89:122003} which 
in combination with updated parton density functions and a combined fixed-order and resummed 
calculation gave an increased 
prediction. New measurements at Run II have also been made by the CDF collaboration which probe 
down to very low transverse momenta. In combination with a measured cross section lower (but 
consistent) than the Run I data, and the above theoretical improvements, the data and theory 
now agree very well. 

The first result from HERA\cite{pl:b467:156} also revealed a large discrepancy with NLO 
QCD predictions. This analysis presented an extrapolated quantity, whereas later 
measurements\cite{hera-b} also presented measured 
quantities. The measurements in 
photoproduction are well described by the prediction from NLO QCD 
and the data from the two collaborations also agree well. The H1 data is somewhat higher than 
that from ZEUS; the difference is concentrated at low $p_T^\mu$ where the H1 data is also above 
the NLO calculation. The measurements in deep inelastic scattering are also generally described 
by NLO QCD although some differences at forward $\eta^\mu$ and low $p_T^\mu$ are observed by 
both collaborations. However, inclusive measurements which lead to a measurement of the beauty 
contribution to the proton structure function\cite{h1-f2b} are well described 
by QCD.

The situation for the QCD description of $b$ production has recently changed significantly. In 
general, QCD provides a good description of the data with some hints (a few sigma) at differences 
in specific 
regions. Certainly, there is no longer a difference of a factor of 2--3 independent of $p_T$. 
The HERA experiments will produce several new measurements in the next few years of higher 
precision and covering a larger kinematic region at both low and high $p_T$ and forward $\eta$. 
Allied with expected calculational and phenomenological improvements, a deep understanding of 
beauty production should be achieved by the turn-on of the LHC.

\section{Charm production}

Due to its larger cross section, more high-precision and detailed measurements of charm 
production at HERA have been made. However, due to length limitations, the reader is referred 
to a previous review\cite{wing-hera-lhc} which discussess open charm production, the contribution 
of charm to the proton structure function and universality of charm fragmentation.

\end{document}